# A simple phenomenologic model for particle transport in space-periodic potentials in underdamped systems

I.G. MARCHENKO[1,2(a,b)], I.I. MARCHENKO[3], A.V. ZHIGLO[1]

[1] NSC "Kharkov Institute of Physics and Technology", Akademicheskaya 1, 61108 Kharkov, Ukraine
[2] Kharkov National University, Svobody Sq. 4, 61077 Kharkov, Ukraine
[3] NTU "Kharkov Polytechnic Institute", Frunze 21, 61145 Kharkov, Ukraine

**Abstract** – We consider the motion of an underdamped Brownian particle in a tilted periodic potential in a wide temperature range. Based on the previous data [1] and the new simulation results we show that the underdamped motion of particles in space-periodic potentials can be considered as the overdamped motion in the velocity space in the effective double-well potential. Simple analytic expressions for the particle mobility and diffusion coefficient have been derived with the use of the presented model. The results of analytical computations match well with numerical simulation data.

Despite particle diffusion and transport in washboard-type potentials have been subject of active research for decades [2,3] investigations in this field keep surprising us. In particular, it was shown in [1] that the particle diffusion coefficient in low friction systems can exponentially rise with a temperature drop. This research is important as it deals with both academic and practical issues. This problem is involved in physical processes that occur in Josephson tunneling junctions, superionic conductors, phase-locked-loop frequency control systems, etc. [2]. The research of the so-called Brownian motors is of special interest [3,4]. Studies of the features of the transport of atoms and point and linear defects in the crystal lattice in external fields are of extreme importance for the development of new technologies for material science. Thorough discussion of applications, references, and methods of studying Brownian diffusion in washboard potentials can be found in [2].

Successful analytical methods were developed for describing the particle diffusion and transport in overdamped problems [5]; however, these methods are of limited usability for systems with low energy dissipation. In this paper we construct a simple phenomenological model of the particle transport in underdamped systems, aiming it in particular at experimentalists. Following our previous work [1] we analyze the particle diffusion and mobility in a wide temperature range by means of numerical simulation. We then propose an analytical model and compare its results with those of computer simulation.

The particle motion is described by Langevin equation. We consider the case of one-dimensional diffusion in a spatially periodic potential $U(x)$

$$m\ddot{x} = -\frac{d}{dx}U(x) - \gamma\dot{x} + F + \xi(t), \qquad (1)$$

under the action of additional constant force $F$. Here $x$ is the particle coordinate, $m$ is its mass, $\gamma$ is the friction coefficient. $\xi(t)$ describes thermal fluctuations, assumed Gaussian white noise,

$$\langle\xi(t)\xi(t')\rangle = 2\gamma kT\delta(t-t'); \qquad (2)$$

here $k$ is the Boltzmann constant, and $T$ is the temperature. Overdot stands for time differentiation.

The potential energy of the particle is

$$U(x) = -\frac{U_0}{2}\cos\left(\frac{2\pi}{a}x\right) \qquad (3)$$

where $a$ is the constant of the one-dimensional lattice. The spatially periodic force exerted upon the particle by the crystal lattice is $F_{lat} = -F_0 \sin(2\pi a^{-1}x)$, where $F_0 = \pi a^{-1}U_0$.

The parameters of the potential are the same as in [1, 6]. The simulation scheme and statistical analysis used here are described in those papers. To compare our results with those obtained by other authors we use dimensionless temperature $T' = kT/U_0$ and friction coefficient $\gamma' = \gamma a(mU_0)^{-1/2}$ [2,7].

In the previous paper [1] the temperature dependence of the diffusion coefficient of an ensemble of particles moving in the tilted periodic potential was studied. In this paper we are more interested in directional transport. The average velocity of the particles is used as a characteristic of this motion,

$$\langle V \rangle = \int_{-\infty}^{\infty} V N(V) dV; \qquad (4)$$

[a] E-mail: `march@kipt.kharkov.ua`
[b] Present address: NSC "Kharkov Institute of Physics and Technology", Akademicheskaya 1, 61108 Kharkov, Ukraine

I. G. Marchenko, I. I. Marchenko, A.V. Zhiglo

here $N(V)$ is the particle distribution function in velocity space.

Particle diffusion and transport in the underdamped case was first studied by computer simulation in [8]. Papers [9−10] present further studies of the $\langle V \rangle(F)$ relation for different $\gamma$'s. The temperature dependence of the mobility was not investigated properly. We study the $\langle V \rangle(T)$ dependence in detail in this paper.

Fig. 1 shows the dependence of mobility $\mu = \langle V \rangle / F$ times the $\gamma$ on the force $F$ for different temperatures. It is known that $\mu$ is asymptotically independent of $F$ at low force values; it satisfies the Einstein relation $\mu = D/(kT)$. $D$ here is the particle diffusion coefficient. In spatially periodic structures $D$ satisfies the Arrhenius relation $D = D_0 \exp(-U_0/kT)$. This linear dependence $\langle V \rangle = \mu F$ is shown as the lower dashed line in the inset in Fig. 1. An upper dashed line in the inset corresponds to $\langle V \rangle = \mu F$ with $\mu = 1/\gamma$; such a regime (of the particle motion in viscous medium) is realized at large $F$. Fig.1 shows that the mobility at an arbitrary force value is between the two asymptotes corresponding to the Einstein relationship for the periodic lattice and for viscous drag. As the temperature drops a more abrupt transition from one functional relationship to another is observed.

zero mobility occurs. This pattern of a hysteresis is often used for the qualitative interpretation of simulation results.

Fig. 1 shows the simulation results for the mobility as a function of the driving force $F$ in the presence of thermal fluctuations. It is seen that an increase in temperature results in $\mu\gamma$ curves becoming flatter. All these curves intersect at one point, at $F \approx 0.095 F_0$ (when $\gamma = 0.02$). This force value corresponds to $F_{02}$ according to classification in [2].

To explain the features of $\langle V \rangle(F,\gamma)$ behavior we consider how the distribution function $N(V)$ changes with an increase in force. It has been known for a while that $N(V)$ is bimodal; we need to understand its shape better, at different $F$. In our previous paper [1] Fig. 3 showed the change in $N(V)$ with an increase in $F$. One maximum corresponds to zero velocity value and the other one to $V = F/\gamma$. Albeit substantial redistribution of the particles between the locked and running solutions takes place with an increase in force the qualitative shape of the curves remains unchanged. As demonstrated in Fig. 2 (in current paper), showing fitting of the data obtained in [1], $N(V)$ is accurately represented as a sum of two Gaussians with the same widths but different amplitudes. The fitting parabolas in Fig. 2 have maxima at velocity values of $V = 0$ and $V = F/\gamma$.

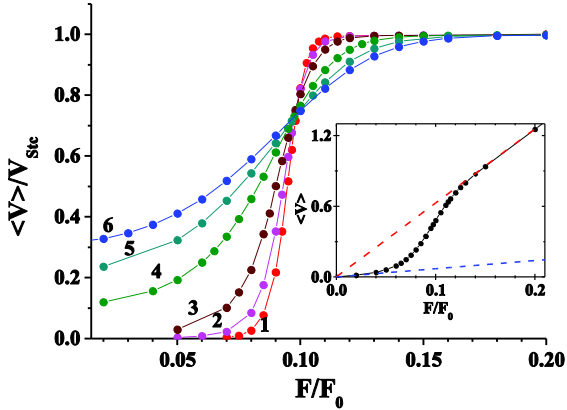

Fig. 1: The dependence of particle velocity $\langle V \rangle$ on the force at different temperatures. $V_{Stc} = F/\gamma$. $V/V_{Stc} = \gamma\mu$. The friction coefficient $\gamma' = 0.141$. $1 - T' = 0.097$, $2 - T' = 0.129$, $3 - T' = 0.194$, $4 - T' = 0.388$, $5 - T' = 0.582$, $6 - T' = 0.776$. Inset: $\langle V \rangle(F)$ at $T' = 0.388$. The Einstein relation for periodic lattice and the stationary velocity of the particle motion in viscous medium are shown by the dashed lines.

A pattern of a particle motion at low friction is described in [2]. When the driving force $F$ adiabatically increases from zero in the absence of thermal fluctuations the particle mobility is equal to zero up to certain force value, $F_{03}$. This is the so-called locked solution. At $F = F_{03}$ the mobility is changed abruptly until the value of $\mu = 1/\gamma$ is reached. At $F \geq F_{03}$ the particle momentum acquired while driven through the lattice period is sufficient to overcome the bonding force at the lattice sites; the particle then moves in the direction of $\vec{F}$, with oscillating velocity. The particle velocity averaged over the oscillation period is $V_{Stc} = F/\gamma$. This is the running solution [2]. If the force is adiabatically reduced from large values, the mobility remains constant, $\mu = 1/\gamma$, down to the value of $F = F_{01}$, at which the reverse jump to

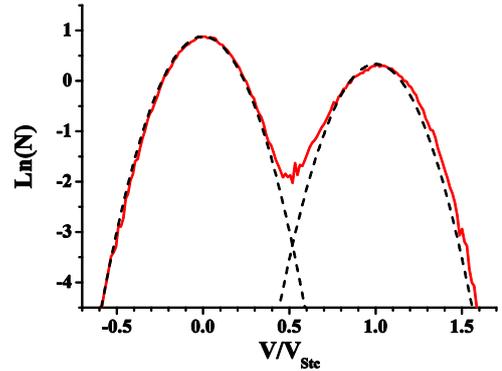

Fig. 2: The velocity distribution function for $F = 0.09 F_0$. $T' = 0.129$, $\gamma' = 0.141$. Approximation with two Gaussians with dispersions $(2kT/m)^{1/2}$ is shown as a dashed line.

Least-square fitting shows that in the entire investigated range of actuating forces $F$ the temperature dependence of $N(V)$ can be rather accurately described by

$$N(V) = A \exp\left(\frac{mV^2}{2kT}\right) + B \exp\left(\frac{m(V - F/\gamma)^2}{2kT}\right) \quad (5)$$

with $F$-dependent $A$ and $B$.

To explain the temperature behavior of the particle transport the qualitative pattern described in [2] requires further development. Let us consider first the case of deterministic particle motion (with thermal noise absent). The particles may have different initial conditions, in particular velocities. At fixed values of $F$ and $\gamma$ different classes of solutions of the equation of motion are realized depending on the value of initial velocity $V_0$. Fig. 3 shows numerical solutions of the equations of motion for different initial conditions. All the particles were put into a local minimum of the total potential energy $U(x) - Fx$ at $t = 0$. Depending on its initial velocity $V_0$ the particle is either in a locked state (curve 1), or evolves towards the running state, moving with the average speed of $\overline{V_F} = F/\gamma$ (curves 2 and 3). It is seen in Fig. 3 that there exists certain critical value of the initial velocity, $V_{cr}$. At $V_0 < V_{cr}$ the particle is in a locked state, and at $V_0 > V_{cr}$ a running solution ensues.

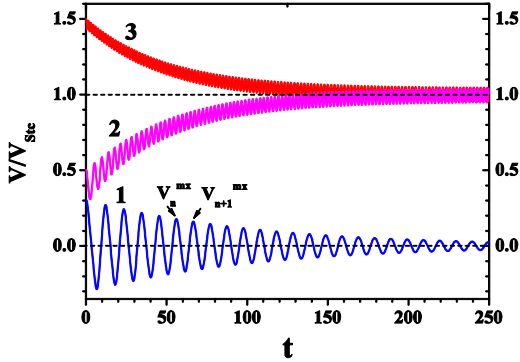

Fig. 3: Time dependence of the particle velocity for different initial velocities $V_0$. $V_{Stc} = F/\gamma$.

The value of $V_{cr}$ depends on $F$ and $\gamma$. Fig. 4 shows the diagram of existence of locked and running solutions depending on the value of the actuating force $F$ at a fixed value of $\gamma' = 0.141$. If $F$ is lower than certain critical value $F_{01}$ the system allows only one, locked, solution. At $F > F_{01}$ two solutions appear, a locked and a running one. Which solution is realized depends on the initial velocity $V_0$. At $V_0 < V_{cr}$ the particle dynamics evolves towards a locked solution and at $V_0 > V_{cr}$ towards a running one. The dashed line in Fig. 4, which separates the solutions, is the relation $V_{cr}(F)$ for the given friction coefficient.

Fig. 5 shows $V_{cr}(F)$ for different values of $\gamma'$. It is seen that $V_{cr}(F)$ varies approximately linearly with the force near its critical value, for all values of $\gamma'$ studied. Analysis of the data of computer simulation also shows that the value of $F_{cr}$ is a linear function of $\gamma'$ at low friction. The simulation data for $V_{cr}(F,\gamma)$ near $F = F_{01}$ are accurately approximated as

$$V_{cr}(F,\gamma) = V_{cr}^0 \left( \alpha + \beta \gamma' - \varepsilon \frac{F}{F_0} \right) \quad (6)$$

with dimensionless fitting parameters

$$\alpha = 0.88, \beta = 0.60, \varepsilon = 2.125. \quad (7)$$

$V_{cr}^0$ is the particle velocity required to overcome the potential barrier when friction is absent: $V_{cr}^0 = (2U_0/m)^{1/2}$.

Let us study the transition to the stationary solution in the absence of thermal noise. Fig. 6 shows a kind of smoothed over oscillations particle acceleration, as a function of the particle velocity. Namely, discrete particle acceleration values $\dot{V}_{n+1/2}^{mx}$ were calculated based on the difference between consecutive maximum velocity values, $\frac{dV_{mx}}{dt} \equiv \dot{V}_{n+1/2}^{mx} = \frac{V_{n+1}^{mx} - V_n^{mx}}{t_{n+1} - t_n}$; the lower index numerates the maxima. One such pair of consecutive $V_n^{mx}$ is pointed with arrows in Fig. 3. The presented curves correspond to different values of the actuating force. Each plot was constructed using the data obtained for different values of the initial velocities of the particle.

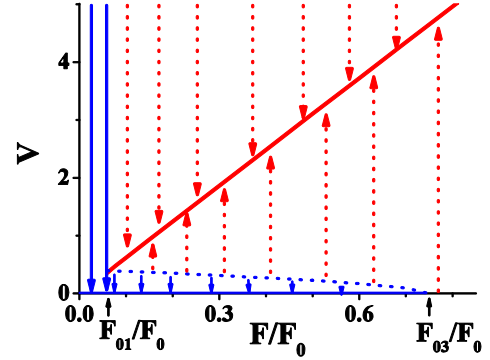

Fig. 4: A phase diagram of locked and running solutions in $(V,F)$ plane. At $F < F_{01}$ only locked solutions exist (the locked phase is colored blue). At $F \geq F_{03}$ only running solutions are realized (colored red). At intermediate $F$ both solutions coexist: at initial $V$ below the critical velocity (shown as the dashed line) the particle is eventually stopped at the full potential minimum; at larger $V$ the particle tends to the running solution, drifting with average velocity $F/\gamma$ (solid red line). $\gamma' = 0.141$.

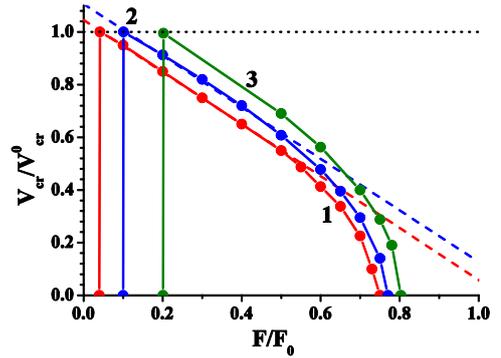

Fig. 5: Dependence of the critical velocity on the applied force in the absence of external noise for different friction coefficients. Linear approximation (6,7) is shown by dashed lines. Curve 1 corresponds to $\gamma' = 0.141$, $2 - \gamma' = 0.354$, $3 - \gamma' = 0.707$. Dotted line - $V_{cr} = V_{cr}^0$.

Fig. 6 shows that the acceleration changes linearly with the velocity, away from the region near the critical velocity. It "pushes" the particle to the stationary solution. At $V < V_{cr}$ the slope $d\dot{V}^{mx}/dV^{mx}$ is $-\gamma/(2m)$ whereas at $V > V_{cr}$ this slope is $-\gamma/m$.

Let us consider the consequences of the contact of the ensemble of particles with a thermal reservoir. While studying the motion of such an ensemble exposed to the action of external force the main attention is usually paid to thermal fluctuations of the force. Classical consideration of this situation is given in Ch. 11 of [2]. Under a contact with the thermal reservoir while the force is changing adiabatically hysteresis loop is formed and one more critical value of force $F = F_{02}$ emerges. At $F = F_{02}$ when the temperature tends to zero a jump from zero mobility to a finite value occurs.

I. G. Marchenko, I. I. Marchenko, A.V. Zhiglo

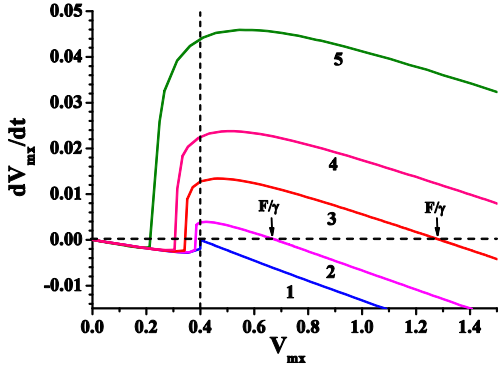

Fig. 6: Particle acceleration as a function of velocity for different values of effective force. The points with zero acceleration are pointed with arrows. No thermal force. 1– $F=0.04F_0$, 2 – $F=0.098F_0$, 3 – $F=0.2F_0$, 4 – $F=0.3F_0$, 5 – $F=0.5F_0$.

For describing average (over oscillation period) dynamics of the particles in low dissipation systems we put forward the following heuristic model, suggested by the $\dot V(V)$ relation noticed above. We assume that particles are in certain effective velocity potential $W(V)$ [11], dependent on the values $\gamma$ and $F$. On the surface this approach is similar to the method in [2]. However, the potential introduced here has a different dimensionality (of $[V]^2/[t]$), as the dynamics is described in velocity space, vs energy space in [2]. This model we present in effect is more akin the model for the motion of active particles for some special type of the potential [11].

The motion of the particles can be described by the following equations:

$$\begin{cases} \dot x = V \\ \dot V = -\dfrac{\partial W(V,F)}{\partial V} + \varsigma(t) \end{cases} \qquad (8)$$

where the noise force satisfies

$$\langle \varsigma(t)\varsigma(t')\rangle = 2Q\delta(t-t') = 2\dfrac{\gamma kT}{m}\delta(t-t') \qquad (9)$$

According to the simulations, the normalized velocity distribution function $N(V)$ is closely approximated by (5). We thus put forward the following form of the effective potential:

$$W(V)=\begin{cases} \dfrac{\gamma}{2m}V^2 + \dfrac{c}{m} & \text{at } V<V_{cr} \\ \dfrac{\gamma}{2m}\left(V-\dfrac{F}{\gamma}\right)^2 + \dfrac{d}{m} & \text{at } V>V_{cr} \end{cases} \qquad (10)$$

where $c$ and $d$ are constants, with appropriate choice of which $N(V)$ must closely match the true distribution function in a wide region in $V$-space around the maxima of $N(V)$.

As it is shown below, an accurate fit for the particle mobility and diffusion coefficients is achieved ($\forall F,\gamma$) when $c$ and $d$ are fixed based on the following recipe. 1). We require the potential continuity at $V_{cr}$,

$$\dfrac{\gamma}{2}V_{cr}^2 + c = \dfrac{\gamma}{2}\left(V_{cr}-\dfrac{F}{\gamma}\right)^2 + d \qquad (11)$$

and 2) we cast the $N(V)$ in the following form

$$N(V)=\begin{cases} \alpha\exp\left(-\dfrac{m}{2kT}V^2 - c\dfrac{m}{\gamma kT}\right) & \text{at } V<V_{cr} \\ 2\alpha\exp\left(-\dfrac{m}{2kT}\left(V-\dfrac{F}{\gamma}\right)^2 - d\dfrac{m}{\gamma kT}\right) & \text{at } V>V_{cr} \end{cases} \qquad (12)$$

as if there is a twofold degeneracy in running states, where $\alpha=\sqrt{m/2\pi kT}$. The 2$^{\text{nd}}$ relation between $c$ and $d$ follows from the normalization

$$\int_{-\infty}^{\infty} N(V)dV = \alpha\int_{-\infty}^{V_{cr}}\exp\left(-\dfrac{m}{2kT}V^2 - c\dfrac{m}{\gamma kT}\right)dV + \\ + 2\alpha\int_{-V_{cr}}^{\infty}\exp\left(-\dfrac{m}{2kT}\left(V-\dfrac{F}{\gamma}\right)^2 - d\dfrac{m}{\gamma kT}\right)dV = 1 \qquad (13)$$

By performing integration this is rewritten as

$$\exp\left(-c\dfrac{m}{\gamma kT}\right)\left[1 + erf\left(\sqrt{\dfrac{m}{2kT}}V_{cr}\right)\right] + 2\exp\left(-d\dfrac{m}{\gamma kT}\right)\times \\ \times\left[1 - erf\left(\sqrt{\dfrac{m}{2kT}}\left(V_{cr}-\dfrac{F}{\gamma}\right)\right)\right] = 1 \qquad (14)$$

Understanding the form of $N(V)$ from the first principles is clearly desirable. This is work in progress.

The average velocities can be found both analytically, through integration (eqs. (4) and (13)), and numerically. The plots of functions $\langle V\rangle$ are given in the lower part of Fig. 7 as a function of the actuating force $F$. Symbols represent the results of numerical integration of Langevin equation (1–2), and solid curves correspond to calculations based on formulae (10–14). The results are shown for two values of the temperature, $T'=0.129$ and $T'=0.388$. It is seen that the simulated results for the average velocity of the underdamped motion of particles in the tilted space-periodic potential agree well with those for an overdamped motion in the double-well velocity potential (10).

If overlapping of the two Gaussians is neglected (by replacing $V_{cr}$ with $\pm\infty$ in Eq. (13)) the computation of the average velocity is considerably simplified. We thus obtain for $\langle V\rangle$

$$\langle V\rangle = \dfrac{F}{\gamma}\dfrac{2}{2+\exp\left(-\dfrac{m}{\gamma kT}F\left(\dfrac{F}{2\gamma}-V_{cr}\right)\right)} \qquad (15)$$

Approximate solutions according to this simplification are shown in Fig. 7 as dashed lines. These also match well the simulation data. This simplification gets less accurate, expectedly, at small $F$, when $F/\gamma$ (and $V_{cr}$) is no longer much smaller than the Gaussian widths, $(2kT/m)^{1/2}$.

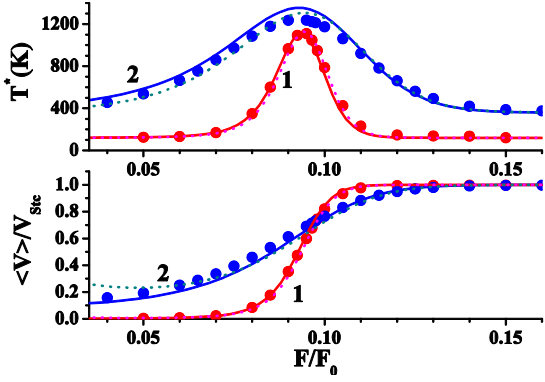

Fig. 7: Dependence of the mobility (lower plot) and the kinetic temperature (upper plot) on the actuating force. Symbols show the simulation results, solid curves denote calculations according to formulae (8−9), the dashed line is approximation (10–11). 1 – $T'$=0.129, 2 – $T'$=0.388. $\gamma'$=0.141.

Fig. 8 shows the agreement between $\langle V \rangle(F)$ curves found in simulations and through solving the model (Eqs. (10−14)) at different values of $T$ and $\gamma$. Curves 1 and 2 correspond to $\gamma'$=0.141, $T'$=0.776 and $T'$=0.194 respectively. Curves 3 and 4 are for $T'$=0.776, $\gamma'$=0.282, $\gamma'$=0.071 respectively. The figure demonstrates that the proposed analytical model accurately describes the behavior of $\langle V \rangle(F)$ for different values of $\gamma$ in the underdamped case.

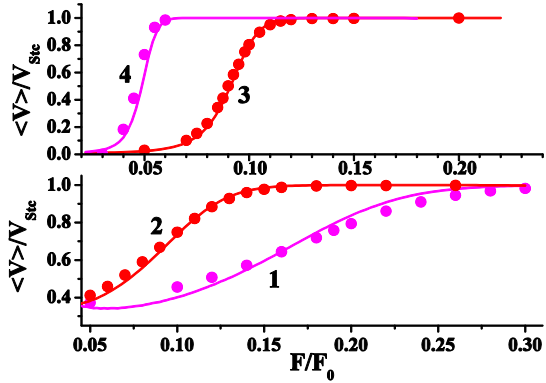

Fig. 8: Average velocity as a function of applied force for different friction coefficients. Curves depict calculations according to Eq. (8−9), symbols show the simulation results. The upper plot – $T'$=0.194, the lower plot – $T'$=0.776. Curve 1 – $\gamma'$=0.282, 2 and 3 – $\gamma'$=0.141, 4 – $\gamma'$=0.071.

Let us consider the way the proposed model describes the temperature behavior of the diffusion. According to the Kubo relation, the diffusion coefficient can be found from the integral of the velocity autocorrelation function, which can be written as the product of the velocity variance $\langle \Delta v^2 \rangle = \langle (v - \langle v \rangle)^2 \rangle$ and its correlation time $\tau_{corl}$ [11]:

$$D = \int_0^\infty d\tau \langle v(t)v(t+\tau) \rangle = \langle \Delta v^2 \rangle \tau_{corl} = \frac{k}{m} T^* \tau_{corl} \qquad (16)$$

Here the kinetic temperature $T^* = \frac{m}{k} \langle \Delta v^2 \rangle$ has been introduced:

$$T^* = \frac{m}{k} \int_{-\infty}^\infty (V - \langle V \rangle)^2 N(V) dV \qquad (17)$$

By numerical integration with the distribution function (12) we find the dependence of kinetic temperature on the acting force. It is plotted in the upper part of Fig. 7, together with the results based on simplified $N(V)$ (that assumes non-overlapping Gaussians for locked and running states; dashed line) and the results for $T^*$ found through direct Monte-Carlo simulation. The simplified expression reads

$$T^* = T + \frac{m}{k} \langle V \rangle \left( \frac{F}{\gamma} - \langle V \rangle \right) = T + \Delta T^* \qquad (18)$$

The Figure demonstrates good agreement of the model results, based on (12), and the simulation results.

Analysis of the obtained results shows that the maximum value of excess in kinetic temperature $\Delta T^*_{mx}$ is approximately equal to $U_0/k$ and is actually independent of the value of $T$. Maximal $\Delta T^*$ is observed at the same fixed value of $F/F_0$, which is also independent of $T$. Thus, at low $T$ values ($T \ll U_0/k$) if the correlation time were fixed the diffusivity should have only changed slightly with the temperature. At the same time paper [1] revealed an exponential increase in $D$ with a temperature drop. This must thus be due to the features of $\tau_{corl}$ behaviour at low temperatures ($Q \ll \Delta W_\pm$); we estimate this below.

For this purpose we use a simple two-state theory in which the velocity performs transitions between two discrete states [12], from $V_-$ to $V_+$ and back, with Kramers rates

$$r_- = \frac{\varpi_{cr} \varpi_-}{2\pi} \exp\left(-\frac{\Delta W_-}{Q}\right) \text{ and } r_+ = \frac{\varpi_{cr} \varpi_+}{2\pi} \exp\left(-\frac{\Delta W_+}{Q}\right).$$

Here $\varpi_-^2$, $\varpi_+^2$ and $\varpi_{cr}^2$ are the absolute values of the curvatures of the potential $W(V)$ at $V=0$, $V=F/\gamma$ and $V=V_{cr}$ respectively.

I. G. Marchenko, I. I. Marchenko, A.V. Zhiglo

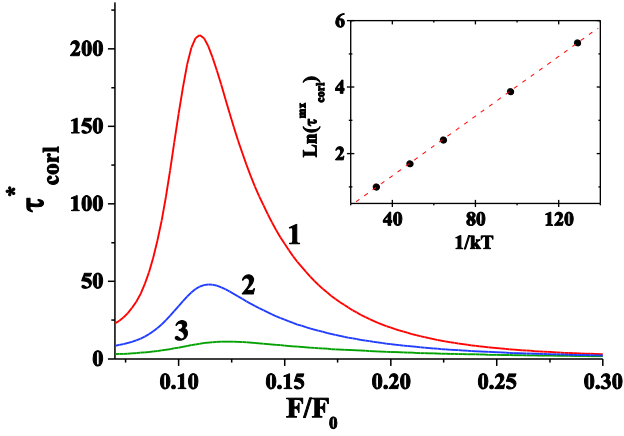

Fig. 9: $\tau^*_{corl}$ as a function of actuating force F for different temperatures. 1 – T = 90 K ($T'$ =0.097), 2 – T = 120 K ($T'$ =0.129), 3 – T = 180 K ($T'$ =0.194). $\gamma'= 0.141$. Inset: temperature dependence of the maximum value $\tau^*_{corl}$.

Approximation (10) for the effective potential is accurate near its maxima, this makes the predictions based on (10) for $D$ and $\langle V \rangle$ correct (as the presented results show). However it is not reliable near $V = V_{cr}$, as can be seen from the absence of continuous derivative of (10) at $V_{cr}$, whereas in part the simulations for $N(V)$ show that the latter is smooth at $V_{cr}$. Thus $\varpi^2_{cr}$ cannot be reliably inferred from our model (10). So we only estimate $\tau_{corl}(F)$ up to an $\varpi_{cr}$-dependent factor (not affecting the main temperature dependence, as is clear below). According to [12] the correlation time is $\tau_{corl} = 1/(r_- + r_+)$. Estimating $\varpi^2_- = \varpi^2_+$ for potential (10) we obtain dimensionless

$$\tau^*_{corl} = \tau_{corl} / \tau_0 = \frac{\exp\left(\frac{mV^2_{cr}}{2kT}\right)}{\left[1+\exp\left(\frac{mF}{2\gamma kT}\left(V_{cr} - \frac{F}{2\gamma}\right)\right)\right]}, \qquad (19)$$

where $\tau_0 = \frac{2\pi}{\varpi_+ \varpi_{cr}}$.

Fig. 9 shows dependence $\tau^*_{corl} = \tau_{corl}(F)/\tau_0$ for different temperatures. It follows from (19) that at the maximum the value of $\tau^*_{corl}$ really behaves as $\tau^*_{corl} \propto \exp(\tilde{T}/T)$ with certain $\tilde{T}$.

The relation (19) obtained in this paper fully coincides with the results of computer simulation in [1], where it was also found that the correlation time increases exponentially with the decrease in $T$. The same functional dependence is observed in results of direct numeric simulations of (8−10).

Summing up, we showed that the velocity distribution function for underdamped particles performing Brownian motion in a tilted periodic potential is closely approximated by the sum of two Gaussians (5), and presented explicit expression (12) for that function valid for a wide range of the temperatures and friction coefficients. Quantities like mobility, diffusion coefficient, kinetic temperature of the particles found in the proposed model agree well with direct numerical simulation results. Proposed distribution function is obtained from the picture of averaged particle dynamics in velocity space (8), with explicitly presented dependence of the velocity potential $W(V)$ on $F$ and $\gamma$. Thus, we showed that the transport of particles in space-periodic potentials in the underdamped case can be treated as overdamped motion of active Brownian particles in velocity space with a double-well potential. This allows one to efficiently use analytical methods developed for the overdamped case. Simple analytic approximations for $\langle V \rangle$, $T^*$, $\tau_{corl}$ can be used in designing, planning and analyzing experiments.

***